# Cybersecurity Attacks in Vehicle-to-Infrastructure (V2I) Applications and their Prevention

**Mhafuzul Islam***
Ph.D. Student
Glenn Department of Civil Engineering, Clemson University
351 Fluor Daniel Engineering Innovation Building, Clemson, SC 29634
Tel: (864) 986-5446, Fax: (864) 656-2670
E-mail: mdmhafi@g.clemson.edu

**Mashrur Chowdhury, Ph.D., P.E., F. ASCE**
**Eugene Douglas Mays Endowed Professor of Transportation**
**Professor of Automotive Engineering, Civil Engineering, and Computer Science**
Glenn Department of Civil Engineering, Clemson University
216 Lowry Hall, Clemson, South Carolina 29634
Tel: (864) 656-3313, Fax: (864) 656-2670
E-mail: mac@clemson.edu

**Hongda Li**
Ph.D. Student
Division of Computer Science
School of Computing, Clemson University
220 McAdams Hall, Clemson, South Carolina 29634
Tel: (864) 986-2018
E-mail: hongdal@clemson.edu

**Hongxin Hu, Ph.D.**
**Assistant Professor**
Division of Computer Science
School of Computing, Clemson University
217 McAdams Hall, Clemson, South Carolina 29634
Tel: (864) 656-2847
E-mail: hongxih@clemson.edu

*Corresponding author
Number of Figures = 4, Tables = 2.
Number of words in abstract = 249, Number of words in text = 5407, Number of references= 31
Word count = 7656 (5407 (words in text) + 249 (words in abstract) + 250*6 (figures and tables) + 500 (references))
Submission date: Nov 15, 2017

## ABSTRACT

A connected vehicle (CV) environment is composed of a diverse data collection, data communication and dissemination, and computing infrastructure systems that are vulnerable to the same cyberattacks as all traditional computing environments. Cyberattacks can jeopardize the expected safety, mobility, energy, and environmental benefits from connected vehicle applications. As cyberattacks can lead to severe traffic incidents, it has become one of the primary concerns in connected vehicle applications. In this paper, we investigate the impact of cyberattacks on the vehicle-to-infrastructure (V2I) network from a V2I application point of view. Then, we develop a novel V2I cybersecurity architecture, named CVGuard, which can detect and prevent cyberattacks on the V2I environment. In designing CVGuard, key challenges, such as scalability, resiliency and future usability were considered. A case study using a distributed denial of service (DDoS) on a V2I application, i.e., the Stop Sign Gap Assist (SSGA) application, shows that CVGuard was effective in mitigating the adverse effects created by a DDoS attack. In our case study, because of the DDoS attack, conflicts between the minor and major road vehicles occurred in an unsignalized intersection, which could have caused potential crashes. A reduction of conflicts between vehicles occurred because CVGuard was in operation. The reduction of conflicts was compared based on the number of conflicts before and after the implementation and operation of the CVGuard security platform. Analysis revealed that the strategies adopted by the CVGuard were successful in reducing the inter-vehicle conflicts by 60% where a DDoS attack compromised the SSGA application at an unsignalized intersection.

## 1.0 INTRODUCTION
The driving force behind a country's economy is a surface transportation system that enables reliable and efficient transportation of passengers and goods. Despite remarkable improvements in vehicle design and performance, which has improved vehicle safety, the number of fatalities is still very high. Annually more than 30,000 fatalities occur on US highways alone (*1*), and in the European Union almost 25,700 road fatalities were reported (*2*). To reduce a large number of roadway crashes and the associated societal costs, different countries have been promoting connectivity between vehicles, known as vehicle-to-vehicle communication or V2V, and between vehicles and transportation infrastructure components, which is vehicle-to-infrastructure communication or V2I. This type of Intelligent Transportation System strategy has the potential to reduce roadway crashes significantly.

V2V and V2I enabled vehicles are emerging as next-generation vehicles for surface transportation systems, leveraging the rapidly growing information and communication technology. Connected vehicles (CVs) share information with other vehicles and with transportation infrastructure using wireless communication that increases traffic safety, provides efficient mobility services and reduces environmental impacts (*3*). However, the risk of cyber-attacks increases as vehicles become more connected through the Internet, and wireless networks. One of the cyberattack gateways to connected vehicles is V2I. Cyberattacks on V2I communication can have devastating consequences if V2I systems are not properly secured. V2I applications present a variety of vulnerabilities that create an attractive target for hackers. For example, hackers could take control of traffic signals, create hazards, and even cause a breakdown of the traffic system. Thus, it is critical to develop new security solutions to protect the V2I environment.

The Connected Vehicle Reference Implementation Architecture (CVRIA) (*4*), as developed by the US Department of Transportation (USDOT), has defined hundreds of CV applications to date. Among them, more than 20 applications are related to V2I. Many of these applications are related to V2I safety (e.g., curve speed warning, in-vehicle signage, and Stop Sign Gap Assist). According to CVRIA, all these applications share some common sub-applications (e.g., speed warnings, intersection safety), and use common processes (e.g., collecting roadside safety data and processing collected vehicle safety data) to support this application layer. For this reason, data flows can pose security risks when shared between these applications. Thus, if a cybersecurity risk is present in any of the shared sub-applications or processes, other applications might be at risk.

Much research has been conducted on enabling privacy, maintaining authentication, and providing integrity by targeting in-vehicle and V2V communication. However, very few studies focus on V2I-level security as a part of the in-vehicle and V2V security solutions (*5–7*). Therefore, research that is solely focused on V2I security solutions needs more attention. Primarily, in this paper, we evaluate the significance of V2I application security. Then, we present an innovative V2I security architecture, CVGuard, which can identify, prevent, and provide countermeasures against security threats and protect V2I applications from being compromised by cyberattacks.

The remainder of the paper is organized as follows: Section 2 provides a literature review followed by the CVGuard development method. Section 3 provides an overview of CVGuard operation and benefits. In Section 4, a case study is conducted by selecting one of the cyberattacks, a DDoS on a V2I application, Stop Sign Gap Assist (SSGA) application. Section 4 also articulates the detection and prevention techniques adopted in CVGuard for a DDoS attack. Moreover, based on the case study on the SSGA application, Section 4 depicts the capability and effectiveness of CVGuard based on its ability to reduce conflicts caused by a DDoS attack. Conclusions, future work, and limitations are addressed in Section 5.

## 2.0 V2I SECURITY-RELATED STUDIES

In this section, the primary goal is to summarize the cybersecurity aspects of the connected vehicle environment. Before developing a cybersecurity solution, it is necessary to determine cybersecurity requirements, different types of potential V2I attacks, and existing V2I security solutions. Also, a review of the emerging technologies is imperative while developing a cybersecurity solution for V2I applications:

### 2.1 V2I Security Requirements

In a connected vehicle environment, vehicles can communicate with internal (i.e., V2S (vehicle-to-sensor)) and external environments, such as, V2V and V2I, including roadside units (RSUs) that use dedicated short-range communication (DSRC) (*8*). An on-board unit (OBU) placed inside a vehicle transmits information to the surrounding environment. RSUs collect data from vehicles and applications installed in an RSU deliver the requested service (*9*). According to CVRIA, all security solutions related to V2I and V2V must focus on three core elements: confidentiality, integrity, and availability (*10*).

*2.1.1 Confidentiality*

The content of the messages exchanged in a V2I environment must be kept confidential, i.e., the content cannot be accessed by unauthorized and unintended users. However, most messages in a connected vehicle environment are public, particularly the exchanged messages between vehicles and infrastructure. Data confidentiality and message confidentiality need to be taken into consideration when designing a secure V2I environment (*11*).

*2.1.2 Integrity*

Messages exchanged between a vehicle and infrastructure must be protected from any unauthorized alteration or modification. This ensures the accuracy, reliability, and trustworthiness of the messages. Every security solution must provide protection from unauthorized intentional or unintentional modifications. Loss of integrity can result in the degradation of services provided by a V2I environment (*12*).

*2.1.3 Availability*

Availability ensures that systems and information are accessible and usable to authorized individuals. In a V2I environment, all RSUs must be available at all times. For example, the critical latency for an intersection collision warning must be less than 100 milliseconds (*13*). Also, all security solutions must provide an operational system even in the presence of faults or risky conditions.

## 2.2 V2I Cyber Attacks

Classifications of malicious activities and security threats as well as attackers in the vehicular environment have been discussed in existing literature. In (*14*) and (*15*), security attackers are classified as internal or external based on membership functionality. Activity levels determine whether the attackers are active or passive. Assessment of intention or reason for an attack has classified attackers as rational or malicious. However, existing work focuses solely on unraveling the problems behind the V2V security attacks. Our research is aimed at V2I cyberattacks in the context of emerging connected vehicle applications. Table 1 shows the cyberattacks that are likely to happen in a V2I interface based on the vulnerability of confidentiality, integrity, and availability with the projected likelihood of an attack ranked as HIGH, MODERATE, and LOW. This table provides an overall understanding of the effect of potential cyberattacks on a V2I interface *(15)*.

*Islam, Chowdhury, Li, and Hu*                                                                                           6**TABLE 1 Different Types of Cyberattacks in V2I Environment** *(5-7, 10, 11, 13–15)*

| Attack Type | Compromised Security Element | | | Effect of Attack on Vehicle-to-Infrastructure environment | Likelihood |
|---|---|---|---|---|---|
| | Availability | Integrity | Confidentiality | | |
| Distributed Denial of Service (DDoS) | HIGH | MODERATE | LOW | Unavailability of service, network collapse, defeating service integrity | HIGH |
| Impersonation | LOW | MODERATE | HIGH | Disturbing the network, hiding identity, and gaining privileges are the primary motives of the attacker. | HIGH |
| Message alternation | MODERATE | HIGH | MODERATE | Affects integrity and confidentiality of the V2I environment, thereby impacting the safety service provided by RSUs. Responding to this type of attack depends on anti-theft solutions provided by vehicles or RSUs. | MODERATE |
| Malware and spam | MODERATE | LOW | HIGH | Malware can cause potential serious disruption in service for an RSU. The impact is considered high due to its long-lasting outages. Malware can be injected into the system during a software update of RSUs. | LOW |
| Eavesdropping | LOW | LOW | HIGH | Both active and passive attackers can steal sensitive and private information, which violates the confidentiality of drivers, and vehicles. | MODERATE |

## 2.3 Emerging Technologies
Currently, several new technologies have emerged to make communication networks more secure, more scalable, and capable of achieving more fine-grained control. In the following sections, we briefly discuss the use of these emerging technologies in terms of V2I security solutions.

### 2.3.1 Edge Computing
Edge computing is a method where the data is processed close to the data source. Edge computing offers dynamic content management, proper resource allocation, and low latency. It ensures high bandwidth by distributing the computational tasks to different edges (*16, 17*). In the context of V2I, security modules need to be near the source of data (i.e., RSU) for faster processing, which can be achieved by edge computing.

### 2.3.2 Software-Defined Networking (SDN)
Traditional network management is complex, brittle, and error-prone (*18*). Those issues are caused by the strong coupling of the data plane (used for data forwarding) and control plane (used for routing logic, packet management, and access control). By decoupling the control plane and data plane, and by introducing programmability, SDN simplifies the network management tasks (e.g., reconfiguring network topology, changing routing logic) and provides more flexibility with respect to data distribution. In a CV environment, SDN can be used to dynamically reroute the data to different destinations as necessary. *(22)* shows an example of leveraging SDN to provide more flexible protection strategies and better resource management in the defense of cyberattacks while minimizing network congestion and user-perceived latency. We believe that SDN can aid in the construction of V2I threat protection systems.

### 2.3.3 Network Functions Virtualization (NFV)
Network functions such as firewalls and intrusion detection systems (IDSs) are currently implemented in specialized hardware such as field-programmable gate arrays (FPGA)(*19*). Although specialized hardware gives better performance, it also leads to high cost, maintenance complexity, and inflexibility. NFV has emerged to overcome these problems(*20*). NFV reduces cost and provides new opportunities in scalability and implementation of network functions in commodity hardware having fine-grained control (*21*). NFV and SDN could play a major role in securing the CV environment in the near future, as leading industry players (e.g., Verizon, AT&T) are embracing these modern technologies (*22, 23*).

This paper focuses on V2I network-level security and presents a newly developed distributed computational platform, CVGuard, which ensures confidentiality, integrity, and availability by leveraging emerging edge computing, SDN, and NFV technologies.

## 3.0 DEVELOPMENT OF CVGUARD
To achieve our research objective, the first step was to define the components and functionalities of the CVGuard system and to create a formulation of the CV environment to detect and prevent networked-vehicle cyberattacks.

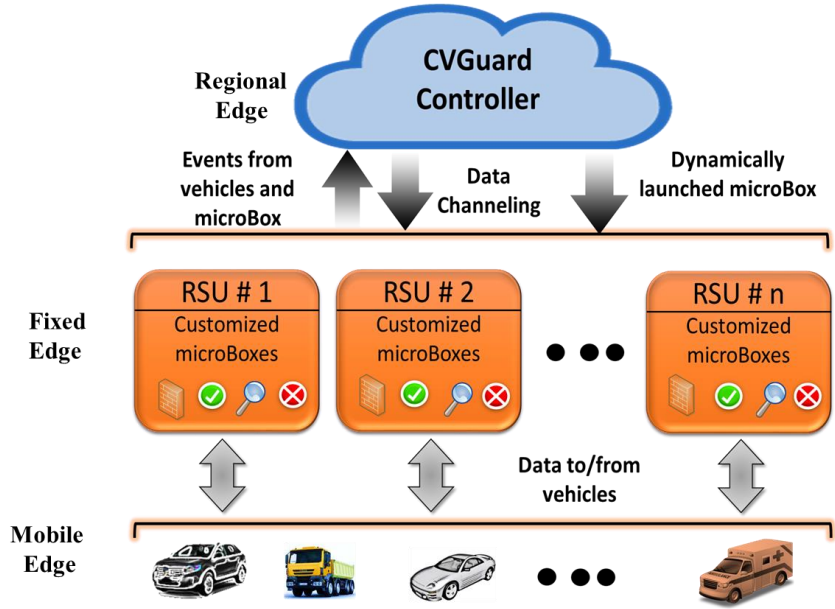

**FIGURE 1 CVGuard: connected vehicle V2I security architecture.**

## 3.1 System Overview

CVGuard is a new software-based security architecture designed to protect V2I applications, leveraging three emerging technologies: edge computing, SDN, and NFV. The primary goal of CVGuard is to detect and isolate any cyberattacks in a V2I environment before they can negatively affect vehicles or transportation networks, which could lead to crashes and impede the adoption of connected vehicle technologies. Figure 1 shows a high-level vision of CVGuard, which contains two major components. The first primary component consists of customized microBoxes (small network-security functions). These small software modules are situated in each fixed edge (e.g., RSU) and act as security gateways for each vehicle (i.e., mobile edge). They also provide dynamic attack capturing and analysis abilities for V2I interfaces. The second major component is the CVGuard controller, which is logically centralized and resides in the cloud as a regional edge where it monitors the contexts of different vehicles, RSUs as well as V2I communications. The CVGuard controller also identifies security threats, analyzes the threats, and controls microBoxes to eliminate the threats. RSUs represent the entry point for potential cyberattacks, and the microBoxes in each RSU do edge computing on incoming data to detect cyberattacks in the first layer of CVGuard. Because of resource constraints, and to provide better service, security functions need to consume as few resources as possible (e.g., memory, processing time). The small software security functions in microBoxes ensure an efficient use of resources in an RSU by being created only when needed and destroyed when no longer necessary. Also, communication networking between RSUs is necessary to disseminate the attack information among neighboring RSUs. By leveraging NFV and SDN technologies, the CVGuard controller can create or destroy security functions flexibly and quickly. The CVGuard controller can also dynamically route attack information and prevention policies to the relevant security functions. Furthermore, the CVGuard controller acts as a centralized security function controller to enable dynamic interoperation and automatic reconfiguration of software security functions, which are distributed in different RSUs.

*Islam, Chowdhury, Li, and Hu* 9CVGuard supports two types of network security functions: i) attack detection as provided by intrusion detection systems (IDS) *(24)* and DDoS detectors *(25)* as well as ii) attack prevention as provided by firewalls and the intrusion prevention system (IPS) *(24)*. Once the attack detection functions discover a potential threat, CVGuard identifies the problem and dynamically launches attack prevention functions to adopt the necessary resolution strategies. Both attack detection and prevention mechanisms in CVGuard are discussed in the following subsections.

*3.1.1 Attack detection*

CVGuard attack detection is based on dynamic security policies and rules, which are developed based on a CV environment where all vehicles are in motion and continuously broadcast the basic safety messages (BSM)–Part 1 at a rate of 10 Hz (10 packets/second) *(26)*. Using wireless communication, each vehicle can communicate within a certain range. From the RSU range of influence, being a static component in the CV environment, vehicles come into the range of communication with the RSU and leave the range after a certain period. To quantify the CV environment, representation of the properties and functionalities of CV components in a proper and realistic manner is essential. For example, each vehicle can be represented as a moving vehicular node having properties, such as speed ($S_i$), and location ($L_i(x,y)$). Table 2 denotes the common properties of the components found in the CV environment:

**TABLE 2 Properties of a Vehicular Node**

| Symbol | Description |
| --- | --- |
| $L_i(x,y)$ | Location of the $i^{th}$ vehicle with a longitude of x and a latitude of y. |
| $L_{rsu}(x,y)$ | Location of the *RSU* with a longitude of x and a latitude of y. |
| $S_i(x,y)$ | Two-dimensional speed of the $i^{th}$ vehicle |
| $S_i$ | Resultant speed of the $i^{th}$ vehicle |
| $R_i$ | Communication range of the $i^{th}$ vehicle |
| $R_{rsu}$ | Communication range of RSU |
| $T_{i,start}$ | Time when the $i^{th}$ vehicle comes into RSU's communication range |
| $T_{i,end}$ | Time when the $i^{th}$ vehicle leaves the RSU's communication range |
| $D_{i,j}$ | Distance between the $i^{th}$ vehicle and the $j^{th}$ vehicle |
| $e_{i,j}$ | Neighbor relationship between the $i^{th}$ vehicle and the $j^{th}$ vehicle |
| $\widetilde{N}$ | Possible maximum capacity of a neighbor's vehicle |
| $N_{rsu}$ | Number of vehicles within the communication range of an RSU |
| $DRR_i$ | Data receiving rate of the $i^{th}$ vehicle |
| $S_{max}$ | Maximum speed of a roadway |
| $S_{min}$ | Minimum speed of a roadway |

Based on the properties of a vehicular node as listed in Table 2, we can formulate the rules needed in a CV environment. The following rules hold true representing the normal behavior of a vehicle without cyberattack:

a) Each RSU has a communication range based on the medium of communication (e.g., the recommended DSRC range is up to 300 m or 984 ft *(27)*). Thus, an RSU will only receive data from those vehicles, which are within the DSRC range. An anomalous location can be detected using the following rule:

$$|L_i(x,y) - L_{rsu}(x,y)| < R_{rsu} \qquad (1)$$

b) The RSU contains the roadway geometry information (e.g., road location, map information) of its surrounding area. Also, the location of each vehicle will be located inside the roadway geometry.

$$L_i(x,y) \in Road\ geometry \qquad (2)$$

c) A predictive approach to the location of vehicles can be computed. If $L_{current}(x,y)$ denotes the current location of a vehicle, and $L_{prev}(x,y)$ denotes the previous location of the same vehicle. Then, the relationship between them can be represented as follows, where $\delta$ is a constant dependent on the traffic situation (e.g., traffic volume, queue):

$$|L_{current}(x,y)| \leq |L_{prev}(x,y) + \delta \qquad (3)$$

d) Usually, each vehicle will maintain a minimum driving distance ($\epsilon$) from other vehicles (Equation 4), and headway ($h_i$) from the vehicle immediately in front (Equation 5).

$$|L_i(x,y) - L_j(x,y)| > \epsilon \qquad (4)$$

$$|L_i(x,y) - L_{i+1}(x,y)| \approx h_i \qquad (5)$$

e) Each vehicle will broadcast BSM at a certain rate, and RSU will receive that data at a rate no higher than its dissemination rate, C1.

$$DRR_i \leq C1 \qquad (6)$$

f) RSU will receive data from each vehicle. To provide a proper service by RSU, the DRR must be higher than the threshold value C2.

$$DRR_i \geq C2 \qquad (7)$$

g) Having a defined communication range, the RSU can accommodate a certain number of vehicles within its communication range.

$$N_{rsu} \leq Capacity\ of\ vehicles\ within\ the\ RSU\ range \qquad (8)$$

h) Vehicles can communicate with a certain number of vehicles inside its communication range.

$$\tilde{N} < Capacity\ of\ neibours\ of\ each\ vehicle \qquad (9)$$

i) The network structure or topology can be determined by defining the neighborhood where two vehicles ($V_i\ and\ V_j$) are defined as neighbors if their physical distance is less than a defined range ($\mu$).

$$V_i\ and\ V_j\ are\ neighbors(e_{i,j} = 1), if\ |L_i(x,y) - L_j(x,y)| < \mu \qquad (10)$$

j) For a certain road section or corridor, the speed of each vehicle will be within a certain boundary:

$$S_{min} \leq S_i \leq S_{max} \qquad (11)$$

k) A vehicle ($i^{th}\ vehicle$) will come into the range of communication with RSU at time $T_{i,start}$ and leave the communication range at $T_{i,end}$. Because of a defined communication range($R_{rsu}$) for the RSU, the difference between $T_{i,end}$ and $T_{i,start}$ will be less than constant, $\tau$.

$$T_{i,end} - T_{i,start} \leq \tau \tag{12}$$

In CVGuard, these rules reside inside the microBoxes that monitor the context of each vehicle. The incoming vehicles' data and a policy set consisting of behavior rules and violation rules are given as input into microBox. Both the vehicles' data and policy set act as a small software security module. The set of policies are dynamically created by the CVGuard controller and fed to the microBoxes. Based on the set of rules and types of violations, a microBox can detect the attack types performed by different attackers. Figure 2 depicts the abstract idea of the detection method. In Figure 2, a cyberattack X type is identified if the vehicular data violates policy P, which consists of rules:{a, b}. Similarly, violation of a policy set Q that consists of a rule set {b, c, d} results in a Y type cyberattack.

*3.1.2 Attack prevention*

Depending on the type of attack, CVGuard can take countermeasures aimed at prevention. On the detection of any attack, the microBox will steer the data from the attackers and legitimate vehicles into the prevention system as shown in Figure 2. Then the attacker's data is quarantined and fed to a CV application based on the need of the application. For example, if a vehicle performs a fake location attack, the RSU can perform an estimation of the correct location of the attackers based on the data collected from other vehicles. Then this localized value can be fed to CV applications. Prevention policies can be changed dynamically based on the type of attack using the CVGuard controller, which makes the system more flexible, robust, and scalable. For example, for a DDoS attack, the violation set includes {e, f}, which creates a higher DRR for one vehicle, causing the DRR of the other vehicles to lower. In prevention of DDoS, CVGuard can drop or filter the data based on the application need to ensure a proper output from the CV application.

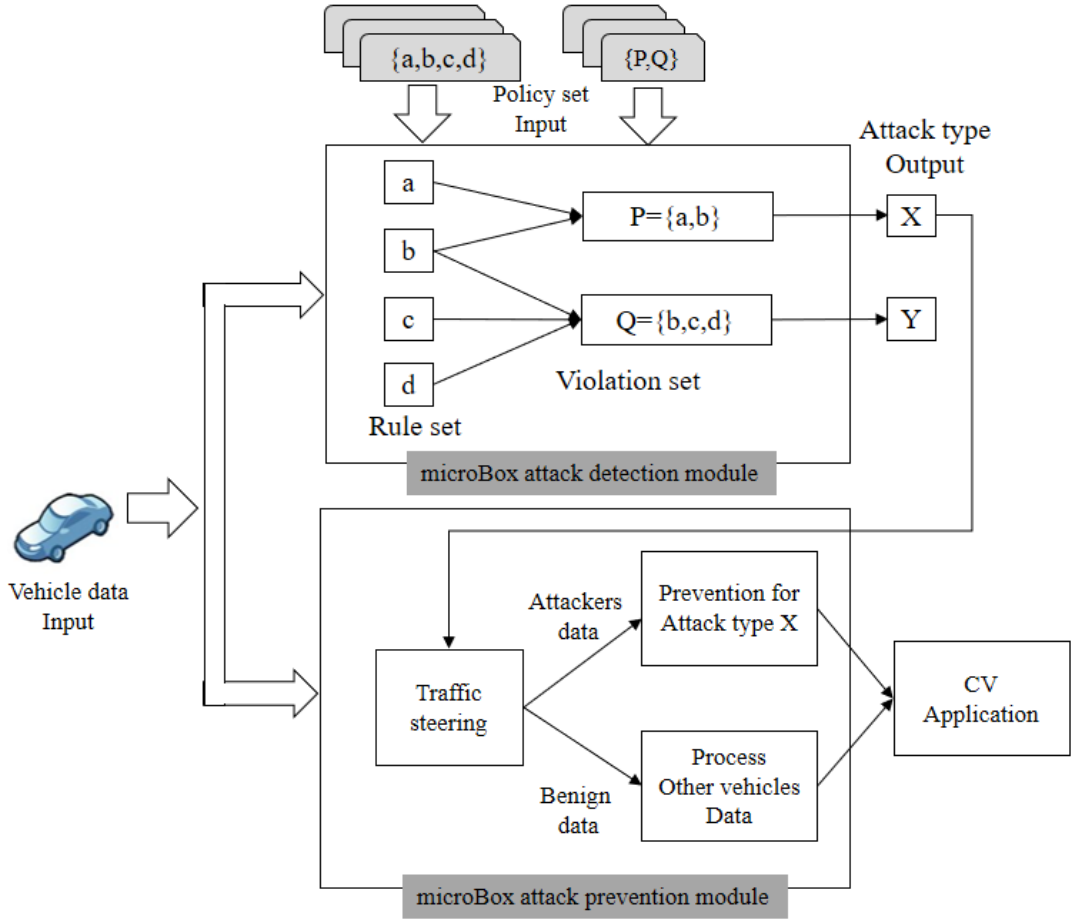

**FIGURE 2 microBox workflow for both attack detection and prevention.**

## 4.0 CASE STUDY: STOP SIGN GAP ASSIST (SSGA) APPLICATION

In this section, we conducted a case study from the vantage point of a V2I application, i.e., an SSGA application. We are going to define and develop a DDoS attack model and later use that model to evaluate the adverse effects on the SSGA application.

### 4.1 SSGA application overview

In our case study, we have selected a V2I safety application, SSGA. The goal of the application is to improve the safety of an unsignalized intersection (having a posted stop sign) by providing safety-related warning and advisory messages to the incoming vehicles at the intersection. In an SSGA application, minor road vehicles are approaching the stop sign, and an RSU is providing safety messages to minor road vehicles based on the state of the intersection (e.g., available safe gap of the major road vehicles). At the intersection, SSGA alerts and warning messages indicate unsafe gaps based on the approaching major road vehicle's speed and distance from the intersection. As a result of real-time warning and alert messaging, the number of right-angle crashes will be reduced in the intersection *(27, 28)*. The performance of the SSGA application depends on the data availability from the vehicles traveling on the major and minor roads. Being a safety-related application, if compromised, the related misinformation can cause crashes at an unsignalized intersection. Because of the safety significance of SSGA, this application is suitable for the case study in determining the effectiveness of the CVGuard system.

### 4.2 Development of DDoS attack model

We focused on a general DDoS (e.g., SYN flood) attack against the CV application running in the roadside infrastructure. The advisory's aim is to exhaust the network and compromise the CV applications. The SYN flood attack is one kind of DDoS attack where a huge number of unnecessary data is transmitted by the attacker (*29*). By creating the SYN flood attack, the services from the CV application can be made unavailable, as this will exhaust the computation resource of the applications. We assume, the advisory model controls its own single or multiple OBUs and can control other vehicles' OBUs to create this attack. The attackers' goal is to compromise safety applications to cause a catastrophic situation. Because a safety application, for example SSGA only takes BSM data to compute the safety operations, we assume that the attacker will create SYN flood with BSM only. For simplicity, we restricted this study to focus on only one element of the roadside infrastructure, RSU.

In the development process of the DDoS attack model, a formulation of a DDoS attack was first developed to evaluate the feasibility of the attack. Second, from commercially available DSRC devices, the value of parameters (e.g., DSRC channel availability, packet size, transmission latency, and capacity) needed for creating the DDoS attack were extracted. Third, using these extracted parameters, in a simulation environment, the attack was launched against the SSGA application. Details of these steps are as follows:

#### *4.2.1 Formulation of the DDoS attack model*

Attacker vehicles launch the DDoS attack by flooding the communication channel. Typically, an attacker uses its maximum transmission capacity to flood the network. To create a breakdown of V2I application, attackers need to transmit more data than the receiver's (e.g., RSU) maximum receiving capacity that can be calculated as

$$Attackers'\ transmission\ capacity \geq Recivers'\ receving\ capacity \qquad (11)$$
$$N_{attackers} * (Tx_a) \geq CRx_{max} \qquad (12)$$

where $N_{attackers}$ is the number of attackers, $Tx_a$ is the attacker's transmission rate, and $CRx_{max}$ is the maximum message receiving capacity of an RSU. Once the attackers flood the network, the RSU will not be able to receive data from legitimate vehicles, and will be unable to provide proper services (e.g., gap alert and warnings) to legitimate vehicles. The ability to perform a successful attack depends on the attackers' transmission rate ($\tilde{x}$), the receivers' receiving capacity (e.g., RSU)($x$), average packet size ($y$), transmission overhead ($\tilde{o}$), and percentage of service channel (SCH) availability ($\rho$). By using these dependent parameters as shown in Equation (12), we can then use Equation (13) to estimate the number of attackers ($N_{attackers}$) needed to create a DDoS attack on a V2I application.

$$N_{attackers} \geq \rho \left( \frac{1}{x} \tilde{x} + \tilde{o} \times \frac{10^6}{8y} \times x \right) \qquad (13)$$

*4.2.2  Parameter extraction from the real devices*

The next step is to verify the feasibility of creating a DDoS attack when the attackers are using real devices based on Equation (13). This feasibility analysis was conducted on commercially available RSUs. The goal was to find the attacker's capacity (e.g., minimum overhead and latency of transmission) with different configurations; then, the extracted values were used to create DDoS attacks in the simulation. Our experiment shows that the minimum transmission overhead at approximately 2.456 milliseconds (ms) with 1 millisecond from the inter-packet gap (IPG) when the device is configured at 3 mbps. Also, on the receiver side, for a single radio, a 46 ms window is available for a SCH in a 100ms time window; thus, the available SCH is 46% of the time of an RSU [16]. Using this latency and other standard configurations as shown below, Equation (13) gives the minimum number of on-board units (OBUs) (min($N_{attackers}$)=3) that are considered as attackers to create a DDoS attack in our experiment, which consists of the following parameters

- Average packet size, $y = 220$ bytes
- Attackers' transmission rate, $\widetilde{x} = 3$ mbps
- Receiver's receiving capacity, $x = 12$ mbps
- Total transmission overhead, $\widetilde{o}$ = IPG+Transmission overhead = (1 + 2.456) ms, So, $\widetilde{o} = 3.456$ ms = 0.003465 sec.
- SCH availability, $\rho = 46\%$
- Minimum number of attackers required, $N_{attackers} \geq 2.83$ or $\min(N_{attackers}) = 3$

## 4.3  CV-Application level impact under DDoS attack

Using the latency parameters from commercially available DSRC devices as described in Section 4.2, different attack scenarios were created to evaluate the impact of DDoS attacks on SSGA in the simulation. With a different number of attackers, the performance of an RSU was evaluated based on the data receiving rate by the application. As the attackers were flooding the RSU, a significant amount of data from legitimate vehicles was lost. As shown in Figure 3, with the increasing number of attackers, DRR by the application decreased due to the DDoS attack. With five attackers, the DRR dropped to 10%, which is the lowest. Because of the data loss from other legitimate vehicles, the CV safety application will be unable to determine the location of legitimate vehicles. This will cause safety application failure to produce accurate output. For example, if the SSGA application produces incorrect output, the conflict between vehicles might have a negative consequence and conflict can potentially lead to a crash. In our experiments, a conflict was recorded as occurring when two vehicles are in close proximity with a time headway of less than 1.5 s (30, 31). In Section 4.6, an assessment was conducted on the probability of a conflict being caused by a DDoS attack.

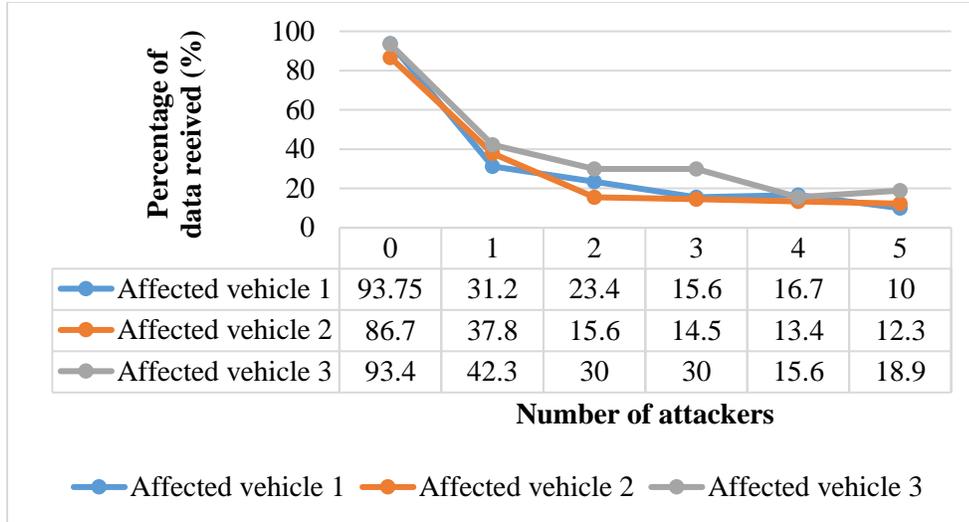

**FIGURE 3 Effect on data receiving rate with different numbers of DDoS attackers.**

### 4.4 Attack detection by CVGuard

As discussed in Section 3, CVGuard monitors the context of vehicles. When rules defined for those vehicles are violated, a malicious activity occurs. CVGuard can detect DDoS attacks and identify the attacker in the CV environment by using the detection policies (e.g., rule set {e, f}) on missing data, and the attacker's DRR. In the case of a DDoS attack, the DRR of the attacker will be high, and the DRR of the other vehicles will decrease as shown in section 4.3. This will cause a violation of policy set {e, f}. According to the rule {e}, RSU will get more data than expected from an attacker. In addition, because of flooding the network, RSU will not receive the data from other vehicles that will cause the violation of rule {f}. Thus, it will get less amount of data from other vehicles to run the application properly thereby making the application service unavailable. However, by monitoring the violation of the policy set:{e, f}, microBox will be able to detect the DDoS attack.

### 4.5 Attack prevention by CVGuard

CVGuard not only identifies the attackers but can also implement countermeasures to mitigate the effect of attacks. In the CV environment, affected components cannot just block the attackers as typically happens in traditional DDoS prevention systems, since in order to perform safety operations, the location information of attackers is also necessary. Upon the detection of the DDoS attack, CVGuard creates a microBox to prevent the DDoS attack. The purpose of this DDoS prevention microBox is to limit the data flow to the safety application. If we can limit the data flow to the application, the application will not be overloaded by unnecessary BSM messages from attackers. We can limit the boundary of message consumption of the application as follows:

$$\alpha \leq DRR_{application} \leq \frac{C}{Number\ of\ vehicles} \times \beta \qquad (13)$$

where, $DRR_{application}$ is the data receiving rate for operating the application properly, $\alpha$ is the minimum data receiving rate required for operating the application and giving a correct output, $C$ is the maximum capacity of the receivers and $\beta$ represents the limiting factor for the upper bound data rate. The values of $\alpha$ and $\beta$ depends on the requirements of the CV applications and the attackers' sending rate. For example, a SSGA application operating at 10 Hz. (*28*) means the receiver should receive data in every 100 ms also can be referred as sampling time. For the SSGA application, the requirement consists of the safe distance and the average velocity of the corridor. The determination of sampling time $T_{interval}$ is crucial in providing a better service to the other vehicles. The DDoS prevention module for SSGA will decide the value of sampling. The selection of $T_{interval}$ can be mathematically expressed as

$$T_{interval} \geq \frac{D_{safe}}{V_{avg}} \quad (14)$$

where $T_{interval}$ is the sampling time for collecting data from legitimate vehicles, $D_{safe}$ is the minimum distance for safe operations that depends on the roadway conditions, and $V_{avg}$ is the average vehicle speed on the subject roadway that is extracted from the data of other vehicles. We chose the minimum of 100 ms and $1/T_{interval}$ as $\alpha$ in Equation (13) to ensure that the SSGA application can operate properly. As shown in Figure 2, the prevention module within a microBox will first classify the data into to two categories, the *benign data* and the attackers' data. The benign data will be processed normally and delivered to the app. The attackers' data will be processed by the prevention component. In the DDoS attack case, the data will be sampled according to Equation (14). The primary actions taken by CVGuard upon the detection of a malicious DDoS attack are described as follows.

(a) **Action 1:** According to the detection result, microBox launches the prevention module for a DDoS attack. The prevention module consists of three components: traffic classification, attack mitigation, and normal processing.
(b) **Action 2:** The attack mitigation component calculates $T_{interval}$ based on the attackers' condition (e.g., number of attackers, attackers' data transmission rate) and other vehicles conditions (e.g., number of vehicles, average speed) by considering α and β
(c) **Action 3:** The mitigation component collects legitimate vehicle data for time $T_{interval}$ and attackers' data from $1/T_{interval}$ time interval; it then feeds these data to the CV application.

## 4.6 Evaluation of CVGuard under DDoS attack

Performance evaluations of the CVGuard system under DDoS attacks were conducted using a SSGA applicaiton with and without CVGuard to evaluate the effectiveness of the CVGuard system. In our experiment, three DDoS attackers (as per Section 4.2) compromised the SSGA application, which caused conflicts at the intersection because the SSGA application was unable to determine the location of both the major and minor road vehicles. In the real-world, the roadway traffic conditions change frequently. Our simulations consisted of 50 major road vehicles and a variable number of minor road vehicles (varying from 5 to 45). Using the minor road vs major road vehicle ratio, we were able to determine the effect of the roadway traffic condition (e.g., number of conflicts) during a DDoS attack. We set $T_{interval}$ at 100 ms as the minimum sampling time with β = 1 in our simulation. As shown in Figure 4(a), using three attackers and each of them transmitting at 500 packets/sec, the mean percentage of conflict was about 28.35% before CVGuard intervention and mean percentage of conflict was approximately 12.65% after the CVGuard deployment. In addition, the rate of conflicts depends on the attackers' message transmission capability (e.g., transmission rate). Figure 4(b) shows that if the attackers' transmission rate increases the probability of conflicts in the intersection increases proportionally. However, after CVGuard integration, the probability of conflicts decreased providing more safety under the DDoS attack scenario with three attackers. For instance, we observed a 46.87% to 18.75% drop in the rate of conflicts in our experiment when each of the attacker's data transmission rate was 1000 packets/sec. Figure 4(c) shows the percentage-reduction of conflicts when CVGuard was deployed under the DDoS attack. After CVGuard integration, the possibility of conflict was reduced by 60% on average.

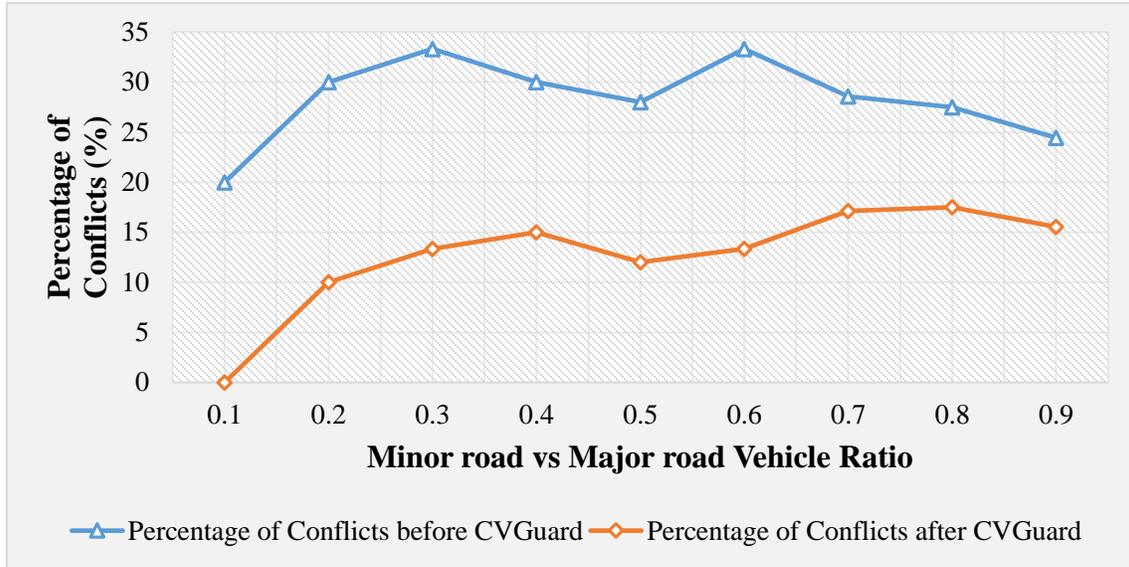

**4 (a) Comparison of percentage of conflicts before and after the deployment of CVGuard having $N_{attackers} = 3$ and Attackers' data transmission rate= 500 packets/sec.**

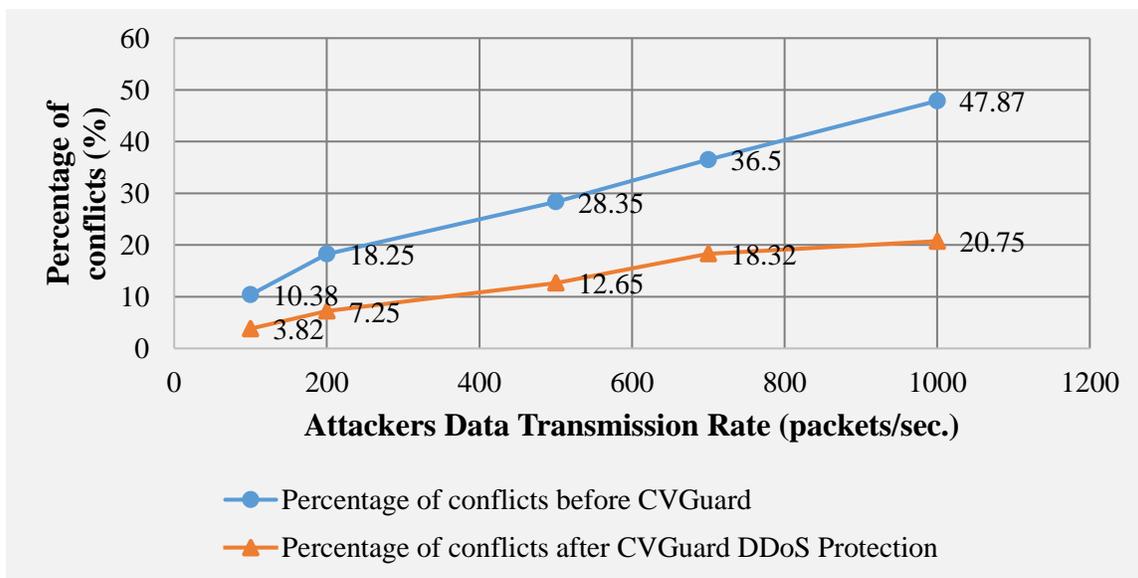

**4 (b) Comparison of percentage of conflicts before and after the deployment of CVGuard having $N_{attackers} = 3$.**

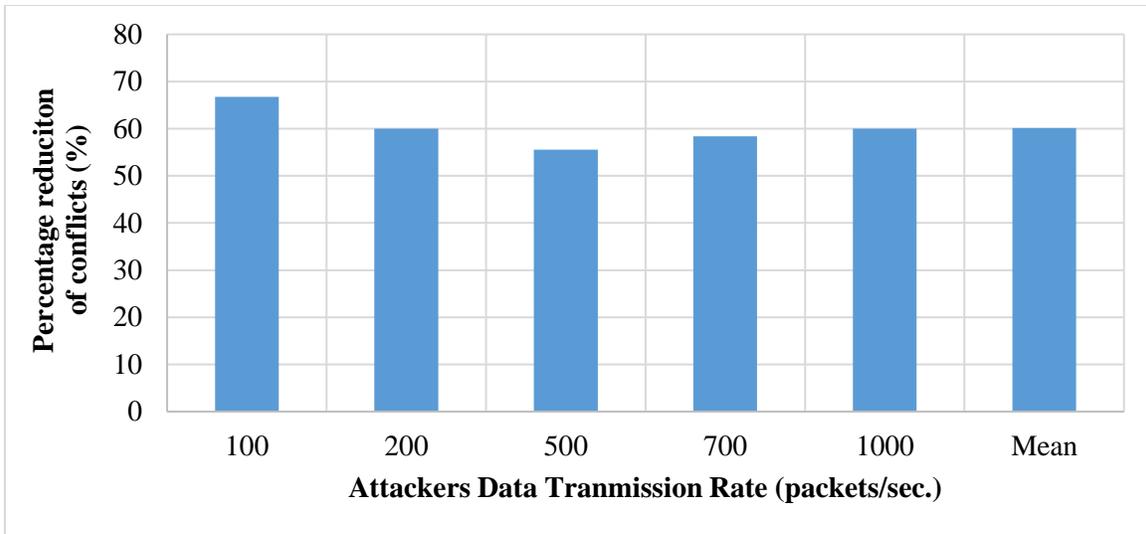

**4 (c) Reduction of conflicts by CVGuard under DDoS attack**
**FIGURE 4 Improvement of conditions before and after the deployment of CVGuard.**

## 5.0 CONCLUSIONS

Because the connected vehicle environment is evolving rapidly, the need for cybersecurity protection is necessary for the near future. Because of cost and maintenance issues, hardware-based security solutions (e.g., hardware security modules) are not feasible. Software-based solutions are much cheaper and dynamically configurable. However, the challenge lies in selecting appropriate methods to protect connected vehicles against cybersecurity threats. Emerging technologies, such as edge computing, SDN, and NFV, have a great potential to make innovations for effective cybersecurity solutions in a connected vehicle environment. This paper presents a novel secure architecture called CVGuard to mitigate V2I cybersecurity threats by leveraging edge computing, SDN, and NFV. First, a case study was conducted encompassing a simulated DDoS attack to determine associated adverse impacts on SSGA due to the attack. Simulation analysis shows that because of the DDoS attack, the DRR by an RSU for SSGA dropped to 10% compared to normal DRR (93%). Second, from a V2I application point of view, the drop in the receiving data rate caused the SSGA application to malfunction and created vehicle conflicts in an unsignalized intersection. Third, following the case study, CVGuard proved its capability by detecting and preventing the DDoS attack and by mitigating its adverse effects. Our analysis shows that the strategies adopted by CVGuard were successful in reducing DDoS attack created vehicle conflicts by 60% where a DDoS attack compromised the SSGA application at an unsignalized intersection. Currently, our study is only limited to DDoS attack. Also, we considered 100% connected vehicles in our simulation, but in real-world traffic, a mixed traffic situation will exist with connected and non-connected vehicles in the traffic stream. Future study will include mitigation of the most common types of cyberattack mentioned in section 2, and creation of different mixed traffic scenarios to evaluate the solutions by CVGuard. We will develop additional modules for the CVGuard system to make the system more robust, scalable, and flexible. Because the CVGuard is a software-based solution, it will be possible to develop different modules by other researchers and developers to contribute to the continuous development of the CVGuard platform. Also, more extensive experiments will be conducted in the near future in the real-world CV environment to evaluate the effectiveness and efficiency of the CVGuard system.

## 6.0 ACKNOWLEDGEMENTS

This paper is based upon research supported by the Southeastern Transportation Center under O/E Grant 2016-1017, and the USDOT Center for Connected Multimodal Mobility (Tier 1 University Transportation Center) grant. Any opinions, findings, and conclusions or recommendations expressed in this material are those of the author(s) and do not necessarily reflect the views of the Southeastern Transportation Center (STC) and the USDOT Center for Connected Multimodal Mobility ($C^2M^2$). The U.S. Government assumes no liability for the contents or use thereof.